\documentclass[fleqn,usenatbib]{mnras}

\usepackage{newtxtext,newtxmath}
\usepackage[T1]{fontenc}

\DeclareRobustCommand{\VAN}[3]{#2}
\let\VANthebibliography\thebibliography
\def\thebibliography{\DeclareRobustCommand{\VAN}[3]{##3}\VANthebibliography}

\usepackage{graphicx}	% Including figure files
\usepackage{amsmath}	% Advanced maths commands
\usepackage{amssymb}	% Extra maths symbols

\usepackage{amsmath,amsfonts,bbm,euscript,hyperref}
\usepackage{xcolor}
\usepackage{color}
\usepackage{hyperref}
\usepackage{multicol}
\usepackage{lipsum}

\hypersetup{colorlinks, citecolor=blue, linkcolor=red, urlcolor=blue}

\usepackage{graphicx,capt-of}

\def\O1{{\cal O}(1)}
\def\at{{\tilde a}}
\def\rt{{\tilde r}}
\def\be{\begin{equation}}
\def\ee{\end{equation}} 
\def\beqa{\begin{eqnarray}}
\def\eeqa{\end{eqnarray}}
\title[On Spin Dependence of the Fundamental Plane]
{On Spin Dependence of the Fundamental Plane of Black Hole Activity}

\author[C. \"Unal \; $\&$ \; A. Loeb]{
Caner \"Unal,$^{1}$\thanks{E-mail: unal@fzu.cz }
Abraham Loeb,$^{2}$\thanks{E-mail: aloeb@cfa.harvard.edu }
\\
$^{1}$ CEICO, Institute of Physics of the Czech Academy of Sciences,
Na Slovance 1999/2, 182 21 Prague, Czechia\\
$^{2}$ Department of Astronomy, Harvard University, 60 Garden St., Cambridge, MA 02138, USA\\
}

\date{Accepted XXX. Received YYY; in original form ZZZ}

\pubyear{2015}

\begin{document}
\label{firstpage}
\pagerange{\pageref{firstpage}--\pageref{lastpage}}
\maketitle

\begin{abstract}
The Fundamental Plane (FP) of Black Hole (BH) Activity in galactic nuclei relates X-ray and radio luminosities to BH mass and accretion rate. However, there is a large scatter exhibited by the data, which motivated us for a new variable. We add BH spin as a new variable and estimate the spin dependence of the jet power and disk luminosity in terms of radio and X-ray luminosities. We assume the Blandford-Znajek process as the main source of the outflow, and find that the jet power depends on BH spin stronger than quadratically at moderate and large spin values. We perform a statistical analysis for 10 AGNs which have sub-Eddington accretion rates and whose spin values are measured independently via the reflection or continuum-fitting methods, and find that the spin-dependent relation describes the data significantly better. This analysis, if supported with more data, could imply not only the spin dependence of the FP relation, but also the Blandford-Znajek process in AGN jets.

\end{abstract}


%%%%%%%%%%%%%%%%%%%%%%%%%%%%%%%%%%%%%%%%%%%%%%%%%%

%%%%%%%%%%%%%%%%% BODY OF PAPER %%%%%%%%%%%%%%%%%%

\section{Introduction}

Although nearly every galaxy contains a supermassive Black Hole (SMBH) at its core \citep{1smbh1gal}, the Active Galactic Nuclei (AGNs) are a small portion of this large SMBH family, and jet producing BHs are a tenth of all AGNs \citep{Begelman:1980vb,Kellermann:1989tq}. Such systems have been shown to be described by an approximate power law relation between BH mass, X-ray and radio luminosity, called Fundamental Plane (FP) of black hole activity   \citep{fp1,fp2,fp3,fp4,kordingfp2006}.  For BHs with outflow and an accretion disk, it is usually considered that X-ray luminosity is often linked with the accretion power (though it also gets contribution from the jet power) and radio luminosity is considered as an indicator of jet power. Hence, the FP relation can equivalently be expressed in terms of BH mass, disk luminosity and jet power. The jet and disk connection in such BH systems result in the mutual scaling of the radio and X-ray (or optical) luminosities \citep{flatspectrum,fp1,fp2,fp3,fp4,kordingfp2006,Plotkin:2011dy,opticalFP}.

The Fundamental Plane relation typically applies for radiatively inefficient AGNs with sub-Eddington accretion rates. However, AGN data shows large scatters relative to the FP relation. Since BH spin is expected to have an important role in both emitted radiation and jets \citep{jetspincorrelation} and expectedly the functional dependencies of the jet and disk power on the spin are different, we suggest that the deviations from the FP solution may result from the BH spin. Therefore, by adding BH spin as a new variable, we predict that the data lies approximately on this new 4-variable relation. %\footnote{One can also add $h/R$ as a new variable to improve the fitting plane relation, $h$ being the thickness of the accretion disk and R is the horizon size of the BH.}.   
In this context, we study the spin dependence of the jet and accretion power and translate the results to the FP quantities, radio and X-ray powers. 

On top of the angular frequency and size of the BH horizon, BH spin controls the location of the inner layers of the accreting matter, which  affects the released energy fraction per mass as radiation. Moreover, as inner layers of accreting matter comes closer to the horizon, the magnetic field around the black hole gets stronger. We assumed Blandford-Znajek (BZ) process \citep{BZoriginal} (see also \citep{jetsAGNsreview} for a recent review)  as the source of the jet luminosity, $L_{jet}$, and it predicts the jet power scales quadratically with BH spin for small values of the spin. Our estimation confirms this scaling and we further find that jet power has a stronger functional dependence on spin at higher values.

We use independent spin measurements from  two well-established methods, namely the X-ray reflection and continuum fitting, to test our prediction \citep{largespincatalog,reynoldsbhspinrev}. We found that the spin modified FP describes data remarkably better, reducing the $\chi^2$ error per degree of freedom from $\sim12.27$ to $\sim2.56$ for 10 AGNs. This result could possibly be improved further by taking into account the variation of the accretion disk thickness and the nonlinearities between $L_{jet}-L_{R}$ and $L_{disk}- L_{X}$.% and may allow us to understand in more depth the universal nature of the jet production and accretion for AGNs with sub-Eddington accretion rates. 
This result, if supported with more data, could imply the existence of the BZ in AGN jets on top of the spin dependence of the FP relation.

\section{The Need For a New Variable in FP} 
Fundamental Plane variables (BH mass, radio and X-ray luminosities) are derived from two main variables : mass and accretion rate. However, the 3 AGNs in the Table \ref{taball}, 3C120, IRAS 00521-7054 and MRK 79, show that these two variables are not enough to explain the data. Namely, these 3 AGNs have masses of around $10^{7.75} M_{\odot}$ and X-ray power of around $ 10^{43-44} \,  {\rm erg \cdot s^{-1}}$ implying these BHs have nearly same M and $\dot m$, but their radio luminosities are different by more than 3 orders of magnitude.  Therefore, it is difficult to explain this variation in the radio spectrum for AGNs by only mass and accretion rate. This fact motivates us to suggest spin as an additional missing variable (this point was also discussed in Ref. \citep{fp1} and recently in \citep{dalyjetspin}).

\section{BH Spin Dependence of Disk Luminosity and Jet Power}

Theoretical results and general relativistic magnetohydrodynamical (GRMHD) simulations,  have demonstrated that the spin energy can be a dominant portion of the jet power \citep{      SPINRLAGN,largespindependence1,largespindependence2,jetspincorrelation}. This is supported by  observations indicating jet power that is orders of magnitude larger than the accretion power (e.g. \citep{jetspincorrelation}). Here we focus on the Blandford-Znajek process which is based on energy extraction from a Kerr BH in the presence of magnetic field set by accretion disk \citep{BZoriginal}. We assume other processes will be subdominant such as wide-angle winds or the Blandford-Payne process \citep{BPoriginal} (recent simulation results show that winds in low accretion rate systems carry a small portion of the outflowing energy \citep{negligiblewind}, while BP may still have some important role \citep{onindepofspin}).

We start by expressing jet luminosity, $L_{jet}$ and  bolometric disk luminosity\footnote{We use disk luminosity and accretion power interchangeably throughout the text.}, $L_{disk}$ in terms of three physical parameters : i) BH mass, $M_{BH}$ (or equivalently, the Eddington luminosity $L_{Edd}$), ii) dimensionless accretion rate, $\dot m \equiv \frac{\dot M}{L_{edd}}$, and iii) the dimensionless BH spin, $\at \equiv \frac{c \,  J}{G M_{BH}^2}$, with J being the angular momentum of the BH,
\begin{eqnarray}
&&  L_{jet} %\propto \; {\dot m}^\gamma \;  M_{BH}^\theta  \; {\cal F}(\at) 
\propto {\dot m}^\gamma \;  L_{Edd}^\theta  \; {\cal F}(\at)  \; ;
 \nonumber\\
&& L_{disk} %\propto  \; {\dot m}^\alpha \;  M_{BH}^\beta  \;  {\cal E}(\at) 
\propto  \; {\dot m}^\kappa \;  L_{Edd}^\beta  \;  {\cal E} (\at) \; ,
 \label{genericdiskjet}
\end{eqnarray}
where  ${\cal F}(\at)$ and  ${\cal E}({\at})$ are the spin dependent functions for the jet power and disk luminosity, respectively. Dimensional arguments set $\theta=\beta=1$. 
%In general the jet power is proportional to total inflowing energy density, hence $\gamma=1$. 
As shown in the next section, the matter inflow to the BH and electric current around the BH are interconnected (setting the external magnetic field around BH), and the jet power is related with the inflowing energy density, which sets $\gamma=1$.  Finally, $\kappa$ is set depending on the accretion state.There are three main regimes for the $\kappa$ parameter : i) in the extremely low accretion rates  ($\dot m\ll 10^{-5}$) $3 \la \kappa  \la 6$, ii) for mildly accreting system ($10^{-5}\la {\dot m} \la 0.1$) $\kappa \simeq 2$, and iii) for large inflow rates (${\dot m}\sim {\cal O}(1)$) $\kappa \simeq 1$.
\\

\subsection{Spin Dependence of the Disk Luminosity}
 The spin dependent term, ${\cal E}({\at})$, shows the radiation conversion efficiency of the accreting matter into BH. The energy released as radiation is some fraction of the difference in the energy of the accreting matter at large radius and at innermost stable circular orbit (ISCO). This radiation efficiency increases as the spin of the BH increases because the innermost stable orbit comes closer to the event horizon. Radiation efficiency  can be expressed by using circular equatorial geodesic equation as (see \citep{refgeodesic}).
\be
{\cal E} (\at) = 1 - \frac{\rt^{3/2} - 2 \rt^{1/2} \pm \at }{\rt^{3/4} \left( \rt^{3/2} - 3 \rt^{1/2} \pm 2 \at \right)^{1/2}} \; \;  \bigg|_{\rt = {\tilde r}_{ISCO}} \;,
\label{eqspinmoddisk}
\ee
where $\rt = r  / GM$. This formula produces familiar results such as ${\cal E} (\at=0) \simeq 0.057$ and ${\cal E} (\at=1) \simeq 0.423$.
\\

\subsection{Spin Dependence of the Jet Power}
Although, isolated BHs are characterized by their mass, electric charge (expectedly small) and spin,  numerous astrophysical phenomena can be observed around them in the presence of accretion disk. We assume Blandford-Znajek (BZ) process is the dominant source of the jet power \footnote{Note that the other mechanisms freeing energy from the BH, such as Blandford-Payne and winds,  exist but in numerous GRMHD simulations, BZ is the main outflow reason.}. This process can be interpreted as the radiative energy outflow from a material with finite resistance moving in external magnetic field.  Assume a piece of material with finite resistance rotates in the magnetic field, the electromotive force induced by this rotating body is ${\cal E} \propto B^2 w^2 R^2$, where $w$ is its angular frequency, $B$ is the external magnetic field, and $R$ is its characteristic size.  A similar expression has been derived in BZ formalism as \citep{BZoriginal,BZderive2}
\beqa
 L_{jet}  &=& \, \int S^{r} \, dA \nonumber\\
 &=& \int   \Omega_A (\Omega_H - \Omega_A) \,   \left( \frac{A_{\phi,\theta}}{\Sigma} \right)^2 \,  \left(r^2+a^2 \right)  \,  \Sigma \, \sin \theta \,  d\theta \, d\phi  \;  \big|_{r=r_H} \, \nonumber\\
\eeqa
 where $S^r$ indicates the radial energy flow ($T^{0r}$ component), $A_i$ indicates the electromagnetic vector potential, comma is a partial derivative with respect to given coordinate, $\Omega_A$ is the angular frequency of the field lines, $\Omega_H$ is the angular frequency of the horizon (defined below) and $dA= \Sigma \, \sin \theta$ with $\Sigma= r^2 + a^2 \cos^2 \theta$.
\be
\Omega_H \equiv \frac{1}{2 G M} \frac{\at}{1+ \sqrt{1-\at^2}}.
\ee
In order to extract energy out, the field line angular frequency should be less than horizon angular frequency, $\Omega_A < \Omega_H$. Assuming the field line velocity is approximately the Keplerian values at regions before the plunging, one can evaluate $\Omega_A \simeq \Omega_{Kepler} \big|_{ r=r_{ISCO}}$. This implies that for $\at \la 0.36$, $\Omega_{Kepler} \ga \Omega_H$, no energy extraction is expected and this value approximately sets the threshold for jet production. For moderate and large spin values, the field line frequency is found approximately as half of the horizon frequency, $\Omega_A \simeq \Omega_H/2$, as in Ref \citep{BZoriginal}.

One can define the magnetic field strength as $\epsilon^{ijk} A_{i,j} = \sqrt{-g} B^k$ so that the pressure and energy density resulting from magnetic field can be expressed as $ \propto B^2$ (without metric modifications on the indices). Given the above relations, the jet luminosity can be expressed as
 \begin{eqnarray}
L_{jet} & \simeq&  \int \frac{w_H^2(\at)}{4} \, (B^r)^2  \, \left( 2 \, {\tilde r}_H (\at)  \right)  \, \sin^3 \theta  \;   \Sigma \; d \theta \, d \phi \nonumber \\
& \simeq&  w_H^2(\at) \, \rt_H^2 \, (GM)^2  \, (B^r)^2  \; {\cal I}(\theta)
\end{eqnarray}  
where ${\cal I}(\theta)$ is an expectedly ${\cal O}(1)$ number which can be also modified by the polar dependence of the magnetic field, $ {\tilde r}_H (\at) = r_H / GM = \left(1 + \sqrt{1-\at^2} \right)$, and $
 w_H(\at) \equiv \frac{\at}{1+\sqrt{1-\at^2}} 
$  is the dimensionless angular frequency of the horizon. Note that $ w_H^2 \, \rt_H^2 =\at^2$.

The system is assumed to be axisymmetric (azimuthal) and stationary. In Ref. \citep{mhdinstability1}, it is shown that the weak magnetic fields develop a strong MHD instability (and further numerically studied in \citep{mhdinstability2}), saturating the field strength near equipartition values  \citep{refmriequipartition}. Hence we have,
\be
\frac{B^2}{8\pi} \equiv  \beta \cdot P_{gas} = \beta \cdot (\rho \, c_s^2) \simeq {\cal O} \left(\rho \,  \frac{\left(L/\mu \right)^2} {r^2} \right) \propto \mu \cdot n \cdot \gamma \cdot \left(L/ \mu \right)^2 \,/ \,  r^2 \; ,
\label{eqmagneticestimate}
\ee
with $\beta$ representing a factor  ${\cal O}(0.1-1)$, $\mu$ is the mass of the individual particle, $L$ is the angular momentum per particle (conserved quantity for geodesics at Kerr background), $n$ is the number density of particles,  $\gamma$ is Lorentz factor.

We employ the continuity equation via the conservation of the particle number density, $J^{\nu} = n \cdot u^{\nu}$, which can be expressed as $J^{\nu}_{\,\, ; \nu} = \frac{1}{\sqrt{-g}} \left( \sqrt{-g} \cdot J^{\nu} \right)_{, \nu}=0$. Due to stationarity and  axisymmetry, partial derivatives of time and azimuthal angle vanish. We also assume that number density is a variable averaged over polar angle. We end up with, 
 \be
 (n \cdot  \Sigma \,  u^r),r=0  \;\;  \Rightarrow  - \frac{{\dot M}}{\mu} \equiv n \cdot \Sigma  \, u^r  \qquad {\rm where} \quad u^{r}=\gamma \cdot v_r  \; ,
 \label{eqnumbercons}
\ee 
where  ${\dot M} = {\dot m}\, L_{edd}$. We estimate the scaled specific angular momentum using the geodesic expression at the equator as \citep{refgeodesic}
\be 
{\cal L} \equiv  \frac{L}{ G \, M \, \mu} =  \frac{\left( \rt^2 \,  \mp \,  2\at \rt^{1/2} \, +\, \at^2 \right)}{\rt^{3/4} \, \left( \rt^{3/2}  \, - \, 3 \rt^{1/2} \, \pm \, 2 \at  \right)^{1/2}} \; ,
\ee
where upper/lower signs are for prograde/retrograde orbits. Note that at large large distances, $(r/GM)\gg1$, we have ${\cal L} \simeq \rt^{1/2}$ which is the standard Keplerian result. Using eqn (\ref{eqmagneticestimate}),we estimate the strength of the magnetic field in the innermost layers of the disk as
\be
B^2_{in} \simeq \frac{ \left({\cal L} \cdot G \, M \right)^2 {\dot M}}{\Sigma \, r^2 \, v_r} \bigg|_{r_{in}} \;.
\ee
Next we consider the magnetic field near the horizon, $B_H$, by assuming that the plunging region, the region between ISCO and event horizon, satisfies nearly vacuum properties. The layers of charged particles near the ISCO form a current loop threading the horizon. The standard dipole magnitude for a loop is 
\be
B_H \simeq \frac{I  \cdot A_{loop} }{r^3} \bigg|_{r_{in}} \; ,
\ee
where $I$ is the current magnitude and $A_{loop}$ is the area of the region the current circulates. One can integrate over various rings that contribute to magnetic dipole moment around the horizon, however since the number density of the disk drops rapidly, the ISCO gives the dominant contribution to the magnetic field around horizon which allows us to represent the disk as a loop for this part of the calculation.

The radiation pressure is assumed to be comparable with the magnetic force exerted on the disk. This gives  $B^2 \; \Sigma \; \propto P \;\Sigma \propto I^2$ yielding
 \be
B_H^2 \propto \frac{ {\cal L}_{in}^2 \; (GM)^{2} \; {\dot m} }{r_{in}^4 \; v_r} \; L_{Edd} \propto \frac{{\cal L}_{in}^2 \;  {\dot m} \; \left(G M \right)^{-2}}{{\cal S}^{7/2}({\at}) \, f(\alpha,\epsilon)} \;  L_{Edd}  \;.
\ee 
Typically the radial velocity is expressed as $v_r \simeq v_{Kepler} \cdot f(\alpha,\epsilon)$,  where $f$ is a coefficient set by gas specific heat ratio, $\epsilon$, and viscosity parameter, $\alpha$, described in Shakura-Sunyaev prescription \citep{alphadiskmodel,Accretiondiskbasics}. We neglect the potential spin dependence of $f$. In the last s, we evaluate the expression near the ISCO, ${\cal S}(\at)   \equiv  r_{ISCO}  \; / \;  G \, M $ 
 \be
  {\cal S}(\at) \equiv \left( 3+ Z_2  \mp [ (3-Z_1)(3+Z_1+2Z_2)]^{1/2} \right)\, ,
  \ee
where $\mp$ indicating the prograde/retrograde rotation respectively, with $Z_1= 1+(1-\at^2)^{1/3} [(1+\at)^{1/3} + (1-\at)^{1/3}]$  and  $Z_2= \left(3\at^2+Z_1^2 \right)^{1/2}$ \; \citep{refgeodesic}.  $S(\at)$ is a monotonically decreasing function of spin. We have ${\cal S} (0)= 6$ ($r_{ISCO}(\at=0)=6GM$) and ${\cal S} (1) = 1$ ($r_{ISCO}(\at=1)=GM$). 

\begin{figure}
%\centering{ 
\includegraphics[width=0.48\textwidth]{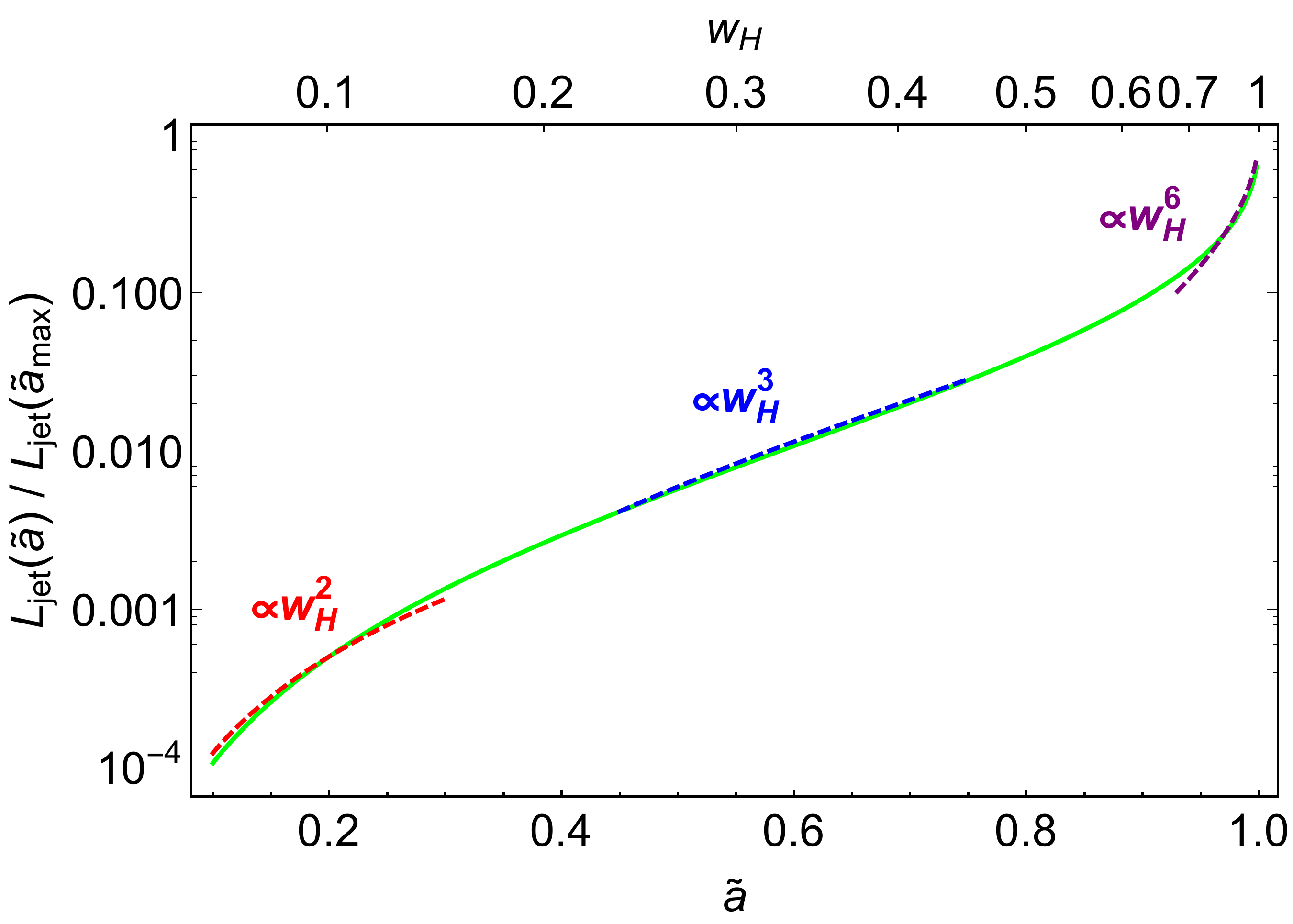}
%}
\caption{The expected nonlinear spin dependence of the jet power in the Blandford-Znajek process.
}
\label{figLjetvsa}
\end{figure}

It is important to note that similar results can be obtained following the arguments in Ref. \citep{flatspectrum} which indicate that the dynamically cooled systems have a magnetic field dependence in the form as ${\dot m}/M$. Furthermore, detailed numerical simulations confirm approximately the radial dependence of the magnetic field estimated above \citep{magneticfieldconfig}. Our jet power expression becomes
\begin{equation}
L_{jet}   \propto \frac{{\dot m} \; {\cal L}_{in}^2 } {{\cal S}^{7/2}(\at) } \; w_H^2 \; {\rt}^2_H \;  L_{Edd} \;.
\label{eqspinmodjet}
\end{equation}

This result shows an interesting behaviour at large values of BH spin. For fixed M and $\dot m$, jet power depends on the amplitude of the angular frequency. At small spin values, $\at \la 0.3$, ${\cal S}$ is nearly constant, hence $L_{jet} \propto w_H^2$. For moderate spin values, we have $L_{jet} \propto w_H^3$, which is valid in the regime $ 0.4 \la \at \la 0.8$, and finally, $L_{jet} \propto w_H^6$, for $\at \ga 0.9$, in the large spin regime. All these regimes are approximately fitted as dashed lines to the exact relation given in Figure  \ref{figLjetvsa}, where $\at$ in lower horizontal axis with a maximum value of $a_{max}=0.998$, and $w_H$ in the upper horizontal axis with a maximum value of $w_{H, max} \simeq 0.939$.   It is remarkable that for geometrically thick and fast spinning BHs, similar results were found in detailed simulations \citep{largespindependence1}.

\section{Evidence for  Spin Modification of the AGN Fundamental Plane %via Measured AGN Spins
}

The Fundamental Plane of black hole activity is a scaling expression for outflowing BHs with sub-Eddington accretion rates relating the BH mass, X-ray  and radio luminosities \citep{fp1,fp2,fp3,fp4}. This relation implies a coupling  between the jet and the disk power, frequently dubbed as jet-disk symbiosis \citep{jetdiskcoup,jetdiskrelation} since X-rays gives a measure of accretion power, and radio emission a measure of jet power. Remarkably, FP expression has been shown to extend from masses of order solar mass up to supermassive BHs, which might be considered as an indication of the universality of jet production.  In a radiatively inefficient systems $L_{disk} \propto L_X \propto {\dot m}^{\alpha} L_{Edd}$ with $\alpha$ being a number in the range 2-3, close to 2.3 \citep{radineff1,radineff2,radineff3} and  $L_R \propto \left({\dot m} \cdot M \right)^{17/12}$ \citep{Blandford:1979za}.
%\citep{fp1,fp2,flatspectrum}.
 Hence, the fundamental plane equation is expressed as \citep{fp1,fp2,fp3}
\be
\log L_{R} =  0.6 \log \left(L_{X} \right)  +  0.78 \log \left(  M_{BH} \right) + constant
\ee
where luminosities are in units of ${\rm erg \cdot s^{-1}}$ and the mass of BH in solar masses, $M_\odot$. 

This result assumes that $L_X$ and $L_R$ are only functions of M and $\dot m$. However for moderate and large spin values, both luminosities also strongly depend on the BH spin. On general grounds, the $L_R - L_{jet}$  relation is quantified as  $L_R  \simeq 10^{40}  \cdot  \left( \frac{L_{jet}}{6 \cdot 10^{43} \, {\rm erg/s}} \right)^{17/12} {\rm erg/s}$ \citep{radiojet,jetradiorelation}. A similar relationship exists between the X-ray power and the bolometric disk luminosity, $  L_X  \propto L_{disk} $.   Since both accretion power and jet power have different spin dependencies, this fact results in many orders of magnitude deviations between  data and the standard FP relation.

  We aim to remove these strong discrepancies by incorporating BH spin into the Fundamental Plane. For that purpose, we express our findings for the jet power  (Eq. \eqref{eqspinmodjet}) and disk luminosity (Eq. \eqref{eqspinmoddisk}) in terms of FP variables, $L_X$ and $L_R$. Finally, we test our predictions for 10 AGNs, given in Table \ref{taball}, whose spin values are measured independently and which are all in sub-Eddington regime.  We label these relations as $SMFP$, the abbreviation of,  ``{\it Spin Modified Fundamental Plane}".
 %
%\beqa
%L_R^{SMFP} &\propto& L_R^{FP} \cdot  [w_H^2 \; \rt_H^2  \; {\cal L}^2 / {\cal S}^{7/2}(\at) ]^{17/12} \; ; \nonumber\\
%L_X^{SMFP}  &\propto&  L_X^{FP}  \cdot {\cal E}(\at) \;.
%\eeqa

The $SMFP$ relation can be expressed as
\be
\log \frac{L_R}{ \left( \frac{w_H^2 \; \rt_H^2 \; {\cal L}^2}{{\cal S}^{7/2}(\at)} \right)^{\frac{17}{12}}}
 = 0.6 \left( \log \frac{{L_X}}{{\cal E}(\at)} \right) +  0.78 \log M + constant
\ee
Introducing the scalings $L_{R,38}= (L_R/10^{38} {\rm erg \cdot s^{-1}})$, $L_{X,40}= (L_X/10^{40} {\rm erg \cdot s^{-1}})$, and $M_{BH, 8}= (M_{BH} / 10^8 M_\odot)$, one can express all the data in the form
\be
\log L^{(i)}_{R,38} =   -0.37  +  0.6 \log \left(L^{(i)}_{X,40} \right)  + 0.78 \log \left( M_{BH, 8} \right)  +  \Delta^{(i)}  \;.
\ee
Superscript $(i)$ is the label of each AGN. In standard Fundamental Plane relation, spin is set to a constant value and spin information is not included. $\Delta$ quantifies the scatter of each data point with respect to the FP predictions as the variance of all the data points is minimized. Hence, by definiton, $\Delta_{FP}=0$ for the standard FP relation  \citep{fp1,fp2,fp3}. However, $\Delta$ changes between $-3$ to $2$ for various data points. 

\begin{figure}
%\centering{ 
\includegraphics[width=0.48\textwidth]{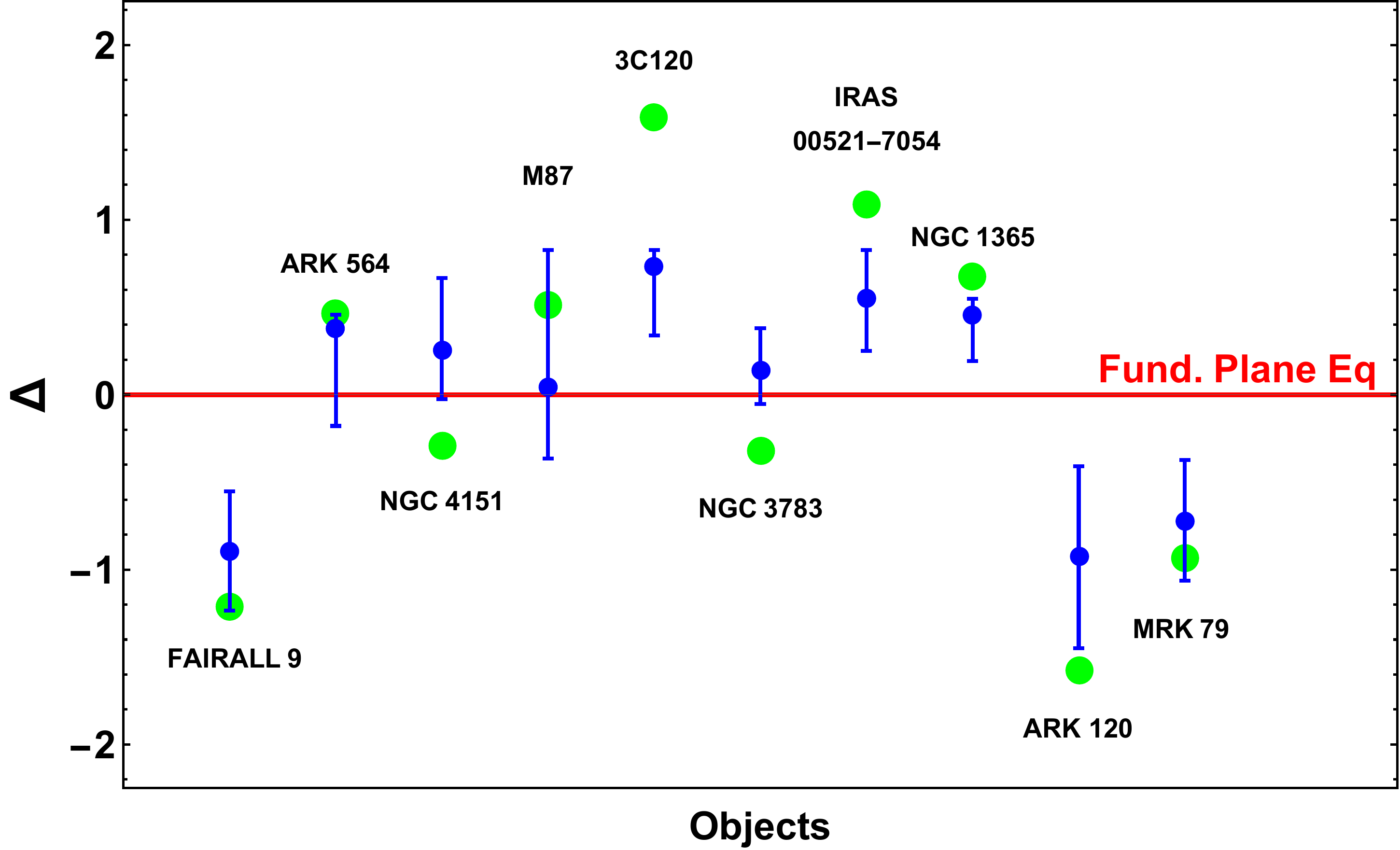}
%}
\caption{The scatter parameter, $\Delta$, is defined in  Eq (\ref{eqdeltadef}) such that the standard Fundamental Plane predicts it to be 0, ie. $\Delta_{FP}=0$. By including BH spin as a new variable, we calculate $\Delta$ including the effect of the spin as in Eq. \eqref{eqdeltaspin}. Data (green points), and Spin Modified Fundamental Plane (SMFP) predictions (blue points) follow the trends and fit the data well.
}
\label{figdelta}
\end{figure}

When BH spin is included, $\Delta$ becomes a function of the BH spin. In Figure \ref{figdelta}, we show that the data fits this new prediction better by studying the suppression of the scatter. In order to show that the inclusion of the spin information decreases considerably the amount of scatter, we invert the equation above and define the scatter function as
\be
\Delta \equiv \log \frac{L_{R,38} }{ 10^{-0.37} \cdot  \left( \frac{ M_{BH}}{10^8 M_\odot} \right)^{0.78} \cdot \left(L_{X,40} \right)^{0.6} }
\label{eqdeltadef}
\ee
In the SMFP, this function is predicted as
\be 
\Delta_{SMFP} \equiv {\cal D}  + \frac{17}{12} \log  \left( w_H^2 \; \rt_H^2 \; {\cal L}^2 \, / {\cal S}^{7/2} \right)  -  0.6 \log \left( {\cal E} \right)
\label{eqdeltaspin}
\ee 
We obtain ${\cal D} \simeq 0.97$, after minimizing the  error variance. We also find that the AGNs nearly on the standard FP prediction corresponds to $\at\simeq 0.9$ \footnote{Note that the spin distribution of heavy BHs are found to be moderate and large \citep{spindistribution} even if the disk is geometrically thick.}. 
Figure \ref{figdelta} shows the data points for the 10 AGNs given in Table \ref{taball} as green dots. The blue dots and corresponding error bars are produced by employing the spin measurements and measurement errors in Eq \eqref{eqdeltaspin}. It is clearly seen that in almost all cases, the spin modified expression tracks the data points better than the spin-blind standard prediction.

\begin{center}
\begin{table*}
%\centering
\begin{tabular}{|c|c|c|c|c|c|c|}  \hline       
Object & \;\; $\at$   \;\; & \;\;  $\log M/M_\odot$ \; \;\;   & \;\; $ \log L_{R} \, / \, {\rm erg \cdot s^{-1}} $  \;\; &  \;\;  $ \log L_{X} \, / \,  {\rm erg \cdot s^{-1}}$   \;\; & \;\; References \;\; \\ \hline 
\;\; FAIRALL 9 \;\; & \;  $0.65 \pm 0.1$ \; & $8.4 \pm 0.12$  & 39.11 & 43.97   & \;\; Br11, Pe04, Wu13, Wu13 \;\;   \\ \hline
\;\;  ARK 564 \;\;  & \;\;  $0.96^{+0.01}_{-0.11}$ \;\; & \;\;  $6.04 \pm 0.13$ \;\;  & $38.59$ &   43.38 &  Wa13, Zh04, Me03, Sa15  \\ \hline
\;\; NGC 4151 \;\;  &  $0.94 \pm 0.05$ & $7.57 \pm 0.2$  & 38.49 &  42.48   & \;\; Ke15, On14, Me03, Wa10 \;\;  \\ \hline
\;\; M87 (NGC 4486) \;\;  & \;\; $0.9 \pm 0.1$ \;\; & $9.81 \pm 0.12$  & 39.85 &  40.46   & \;\;Ta20, Ak19, Me03, Me03 \;\;  \\ \hline
\;\; 3C 120 \;\;  & \;\; $0.994^{+0.004}_{-0.04}$\;\; & $7.74 \pm 0.17$  & 41.36 &  44.06   & \;\;Lo13, Pe04, Ch09, Ch09 \;\;  \\ \hline
\;\; NGC 3783 \;\;  & \;\; $0.92 \pm 0.04$\;\; & $7.47 \pm 0.10$  & 38.78 &  43.10   & \;\; Br13, Pe04, Be18, Be15 \;\;  \\ \hline
\;\; IRAS 00521-7054 \;\;  & \;\; $0.98^{+0.018}_{-0.04}$\;\; & $7.7 \pm 0.18$  & 40.64 &  43.60   & \;\; Wa19, Wa19, Me10, Ta12 \;\;  \\ \hline
\;\; NGC 1365 \;\;  & \;\; $0.97^{+0.01}_{-0.04}$\;\; & $6.7 \pm 0.2$  & 37.65 &  40.60   & \;\; Ri13, Fa19, Me10, Me03 \;\;  \\ \hline
\;\; ARK 120 \;\;  & \;\; $0.64 \pm 0.15 $\;\; & $8.18 \pm 0.2$  & 38.56 &  43.95   & \;\; Wa13, Pe04, Wu13, Wu13\;\;  \\ \hline
\;\; MRK 79 \;\;  & \;\; $0.7 \pm 0.1 $\;\; & $7.72 \pm 0.2$  & 38.35 &  43.12   & \;\; Ga11, Pe04, Wu13, Wu13 \;\;  \\ \hline
\end{tabular}
\caption{The mass, X-ray luminosity [$2-10$ keV], radio luminosity [$\sim 5$ GHz] (the measurement frequencies may vary, typically in the range [$1-100$]GHz) and spin data of 10 AGNs. Doppler boosting effects are neglected. Coding of references are given as follows Br11=\citep{tableref2}, Pe04=\citep{tableref17},
Wu13=\citep{tableref1}, Wa13=\citep{tableref18}, Zh04=\citep{tableref19}, Me03=\citep{fp1}, Sa15=\citep{tableref3}, Ke15=\citep{tableref15}, On14=\citep{tableref4}, Wa10=\citep{tableref5}, Ta20=\citep{m87spin2}, Ak19=\citep{m87mass}, Lo13=\citep{tableref16}, Ch09=\citep{3c120data}, Br13=\citep{tableref6}, Be18=\citep{tableref7}, Be15=\citep{tableref8}, Wa19=\citep{tableref9}, Me10=\citep{tableref10}, Ta12=\citep{tableref11}, Ri13=\citep{tableref12}, Fa19=\citep{tableref13}, Ga11=\citep{tableref14}
} 
\label{taball}
\end{table*}
\end{center}

\section{\bf Analysis with BH Spin in the Fundamental Plane}

We employ the data given in Table \ref{taball},  and compute two quantities that signify the importance of the spin dependence. First, we use the (face value of the) independently measured spin values and calculate standard deviation for $\Delta$ in both FP and Spin Modified FP (SMFP). Second, we compute $\chi^2$ error per degree of freedom, by taking into account the experimental error in spin measurements via using a top-hat probability distribution function for each spin value within a 1-$\sigma$ error range.  In both cases, $N=10$ for our analysis.
\be
\sigma_\Delta = \sqrt{ \frac{\sigma^2}{\# \, {\rm pts}}} =  \sqrt{\frac{1}{N}  \sum_{i=1}^{N} \;  \left( \Delta_{data}^{(i)}- \Delta_{th}^{(i)} \right)^2 } =  \sqrt{\frac{1}{N}  \sum_{i=1}^{N} \;  \left( \delta^{(i)} \right)^2}  \;.
\ee
where "\#pts" indicate the number of sample points. As discussed in the previous section, $\Delta_{th}=0$ for FP , and for Spin Modified FP, it is given in Eq. \eqref{eqdeltaspin}. We obtain  $\sigma_\Delta(FP) =0.98$ and  $\sigma_\Delta(SMFP) =0.49$.  It should be noted that for AGNs with spin information, the variance $\sigma_\Delta(FP)\sim 1$ is also consistent with the larger samples used in Refs \citep{fp1,fp3}  \footnote{Similar scatters were found in Ref. \citep {fp1} ($\sigma_\Delta \simeq 0.89$), Ref. \citep{fp4} ($\sigma_\Delta \simeq 1$), and for LINERS \citep{linersfp} ($\sigma_\Delta \simeq 0.73$).}.

We suggest that the nonlinear effects of spin modification on the radiation efficiency and jet power is the main cause of this behaviour and our SMFP prediction shows a remarkable consistency with the data. Note that the selection of data might have a role at some degree and an analysis with higher precision spin and radiation data could allow us to test our predictions better.

We calculate $\chi^2$ parameter per degree of freedom as
\begin{eqnarray}
 {\hat \chi}^2   \!\!\!\! &=&  \!\!\!\!\!\! \frac{1}{N}  \left( - \log \bigg[ \Pi_{i=1}^{N}\int_{\at_{i,min}}^{\at_{i,max}} d\at_i \, P(\at_i) \exp \left[- \left(\delta^{(i)} \right)^2 / \sigma_{meas,i}^2 \right] \bigg] \right) \nonumber\\
 &=&   \!\!\!\!\!\! \frac{1}{N}  \left(  \sum_{i=1}^{N} - \log\bigg[ \int_{\at_{i,min}}^{\at_{i,max}} \frac{d\at_i}{\Delta \at_i} \exp \left[- \left(\delta^{(i)}\right)^{2} / \sigma_{meas,i}^2 \right] \bigg] \right)  .
\end{eqnarray} 
Here $\sigma_{meas,i}^2 = \Delta L_R^2 + 0.6^2 \Delta L_X^2 + 0.78^2 \Delta M^2$, and $\Delta \at_i= \at_{i,max}- \at_{i,min}$, and $P(\at)=\frac{1}{\Delta \at_i}$ due to top hat distribution estimation.  Nearly all measurements have mass, radio and X-ray measurements with an error of 0.2 in log base, so we set $\Delta L_R = \Delta L_X = \Delta M = 0.2$. We obtain ${\hat \chi}_{FP}^2=12.27$ and ${\hat \chi}_{SMFP}^2=2.56$. Hence the inclusion of spin correction improves the fitting considerably \footnote{This result could also be strengthened by using a Bayesian approach such as one conducted in Ref. \citep{Plotkin:2011dy}.  Doppler beaming effects could be reason of slight outlying of AGN 3C120. Lorentz factor of about 5 and small viewing angle  could improve the $\hat \chi^2$ even further.}. 
  
\begin{figure}
\includegraphics[width=0.48\textwidth]{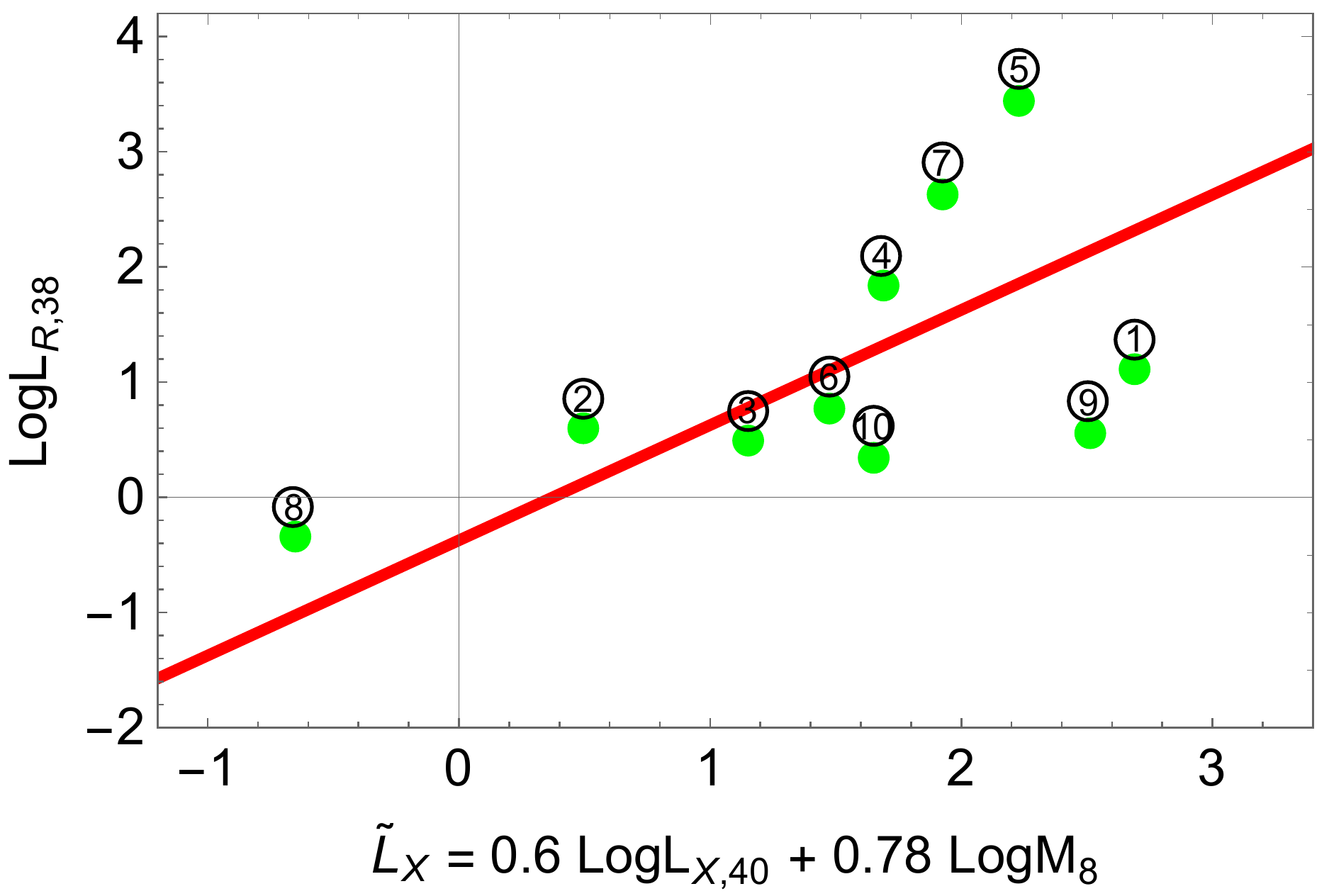}
\includegraphics[width=0.48\textwidth]{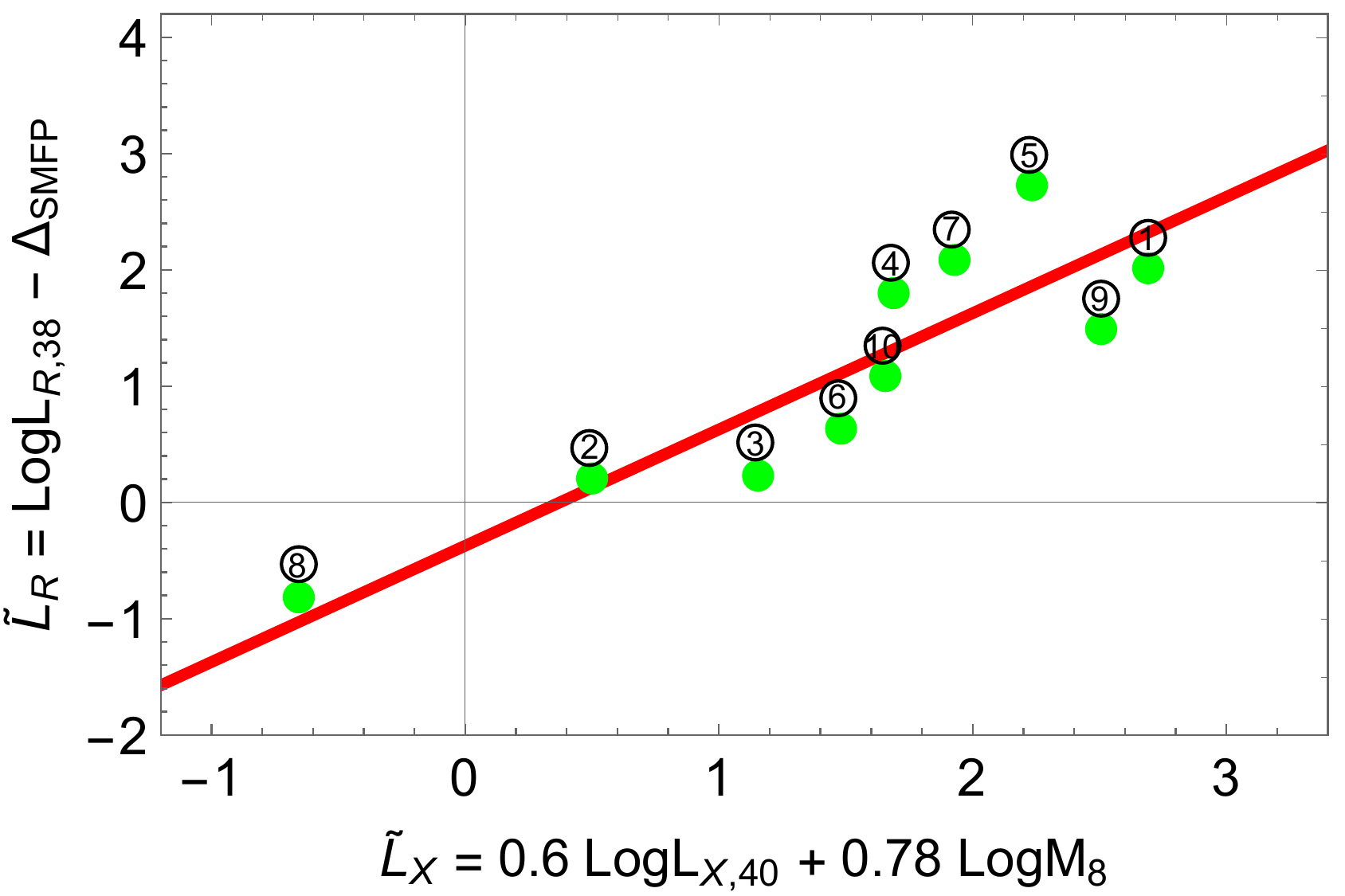}
\caption{Top: The standard Fundamental plane, Bottom: Modified Fundamental Plane with spin dependence.
}
\label{figFPwithwoutspin}
\end{figure}

We define ${\tilde L}_R \equiv L_R - \Delta_{SMFP}(\at)$ and plot in Figure \ref{figFPwithwoutspin}, this new spin modified variable, ${\tilde L}_R$ (on the bottom panel) and $L_R$ (on the left panel)  as a function of ${\tilde L}_X= 0.6 \log L_{X, 40} +0.78 \log M_8$. In both cases the slope of the line is 1. For the standard FP plane, there is considerable scatter (shown in the top panel), but  the scatter decreases considerably with this new variable. However, the prediction curve on the right panel is still not razor-thin. There are potentially some reasons for that including the thickness of the accretion disk enters as an independent variable, data uncertainties, $L_X-L_{disk}$ and $L_R-L_{jet}$ connection, etc and they are discussed in more detail in the next subsections.

\subsection{Spin Dependence of the Fundamental Plane}

 As discussed in the previous section, there exists a natural cutoff value for the spin value in the Blandford-Znajek(BZ) process, $\at\sim 0.36$. In the regime of low spins, winds and Blandford-Payne type processes dominate over BZ. This leads to a nearly constant and small outflow for small spins. However, we estimate that BZ quickly starts dominating the power outflow at moderate spin values and the jet power grows even more nonlinearly for large spins.  At low spin values, the ISCO is nearly constant and the angular frequency grows quadratically. As the spin grows the location of the inner layers of the accretion disk (equivalently the location of the source for the magnetic field) gets closer to the horizon. As a result, the BZ output changes more dramatically with growing spin.  Since the jet and radiation efficiency have different functional dependence on the spin value, this results in  strong scatter of the data from the standard FP relation when the spin information is not included $\sigma_\Delta(FP) =0.98$, and ${\hat \chi}_{FP}^2=12.27$.

If one assumes only quadratic dependence to spin value, the jet power grows from $\at \sim 0.4$ to $\at \sim 1$ about a factor of 6,  and interestingly, the radiation efficiency also grows  by a factor of 6 from small spin values to large spin values, This can be easily seen either from Eq \eqref{eqspinmoddisk} or approximately from the variation of the location of the ISCO in inverse gravitational potential, $(G_N \, M /r_{ISCO})$. Transfering these results to the FP variables, we get that the spin dependent contribution to the X-ray and radio power is modest. When we repeat our statistical analysis for the $L_{jet} \propto a^2$, we find only slight improvement in deviations, namely $\sigma \big|_{L_{jet} \, \propto \, \at^2} \simeq 0.89$ and ${\hat \chi}^2 \big|_{L_{jet} \, \propto \, \at^2} \simeq 9$.
However, the stronger spin dependence of the jet power results in  $\sigma_\Delta(SMFP) =0.49$ and ${\hat \chi}_{SMFP}^2=2.56$.

\subsection{Residual Scatter}
Here we note some potential reasons about the residual scatter in SMFP:
\begin{itemize}
\item $L_{bol} \neq {\rm const} \cdot L_X$ and $L_{jet} \neq {\rm const} \cdot L_R^{12/17}$.

We derived our results for the spin dependence of the jet power and bolometric disk luminosity, then transfered them to the standard definition of the Fundamental Plane relation which employs X-ray luminosity and radio luminosities.  Although there is a strong correlation between $L_X$ and $L_{disk}$, also between $L_R$ and $L_{jet}$, there might be slight deviations from the expressions adopted above \citep{Sikora:2006xz}. 

\item The ratio between the thickness of the accretion disk to the horizon size, $h/r_H$ enters in the expressions for both jet and disk luminosity. Throughout this analysis we implicitly assumed this ratio is nearly same for all the BH systems and this certainly introduces some fluctuations. However, these fluctuations could typically introduce scatter up to an order of magnitude which is subdominant with respect to spin dependence and could be reason for persistent small deviations in the spin modified version of the FP. 

\item There are observational uncertainties in the measurements of the spin, X-ray and radio luminosities. These uncertainties are again up to a factor of a few. Also it is known that jet power can also contribute X-ray luminosity but in the systems we focus on this effect is expected to be subdominant.

%\item The theoretical predictions for the jet power and disk radiation can be sharpened.
\end{itemize}

\section{Summary and Conclusions}
\label{sec:conclusion}
In accreting and outflowing BHs, the Fundamental Plane relation is one of the most important scalings relating three variables: the radio luminosity (indicating jet power), X-ray luminosity(indicating bolometric disk luminosity) and the BH mass. Although standard FP equation gives a good description of jet producing systems with sub-Eddington accretion rate, there are still considerable deviations around this relation. In order to explain these deviations, we investigated  the spin dependence of the jet power  and accretion power.  We can summarize the effects of the BH spin on the jet and accretion power as follows: As the spin grows i) a larger fraction of the gravitational energy can be released as radiation since the inner orbits of the accreting matter could come closer to the horizon, ii) the magnetic field amplitude around the horizon grows since the source for the magnetic field comes closer, iii) the angular frequency of the horizon grows, iv) the size of the horizon decreases.
\\
\\
We derived three main results: 
\begin{itemize}
\item  We estimated the spin dependence of the dominant jet production mechanism, Blandford-Znajek process.  We found that although for small spins, jet power is proportional to angular frequency quadratically as in the famous perturbative result, for moderate and large spins this is not the case. The jet power depends on the angular frequency more strongly (with sixth power) for very large values of spin parameter;

\item  We showed that standard Fundamental Plane cannot explain the data purely by assuming $L_X$ and $L_R$ are only functions of M and $\dot m$. We gave an explicit example by using the data of 3 AGNs which have nearly same mass and X-ray power, but significantly different radio power;

\item  By using data on 10 AGNs, we showed that BH spin could have an important role in the black hole activity. Our Spin Modified Fundamental Plane (SMFP) relation shows significantly well agreement with the data. The many orders of magnitude scatter of the data in the standard FP relation drops to about an order of magnitude scatter in the SMFP. The remaining deviations can be explained by various sources, including the thickness of the accretion disk, uncertainties in the data, the environment around the AGN and uncertainties in the correlation between radio-jet power and X-ray-accretion power.  In conclusion our results, if confirmed with more data, do not only stress the vital role of BH spin in accreting and outflowing BHs, but could also provide a strong observational indication for the existence of the Blandford-Znajek process.

\end{itemize}

\section*{Acknowledgements}

We would like to thank Ruth Daly, Kaz{{\i}}m Yavuz Ek{{\c{s}}}i, Charles Gammie, Gabriele Ghisellini, Kianusch Mehrgan,  Coleman Miller,  Roberto Oliveri, Gizem {{\c{S}}}eng\"or, Bayram Tekin and Federico Urban for discussions on a variety of topics; Tansu Daylan and St\' ephane Ili\'c related with the statistical analysis;  Ivan Almeida, Heino Falcke, Kayhan G\"ultekin, Sebastian Heinz,  Elmar K\"ording, Sera Markoff, Richard Plotkin and Gustavo Soares for their comments on the manuscript. We also would like to thank anonymous referee for her/his comments. C\"U  thanks \"Omer and Hamiyet \"Unal for their support and encouragement throughout this work, and Jeroen Wienk for hosting and helping him in Nijmegen in the chaotic days of the COVID-19 spread. C\"U thanks Harvard University, Munich Institute for Astro and Particle Physics (MIAPP) and the organizers and participants of the workshop Precision Gravity: From the LHC to LISA, International Center for Theoretical Physics (ICTP), Galileo Galilei Institute for Theoretical Physics (GGI), Middle East Technical University (METU) and University of Geneva for their kind hospitality during the progress of this work, and acknowledges the hospitality of the DS\.I G\"olk\"oy E\u{g}itim Tesisleri personnel. C\"U  was supported by European Structural and Investment Funds and the Czech Ministry of Education, Youth and Sports (Project CoGraDS - CZ.$02.1.01/0.0/0.0/15\_003/0000437$) and partially supported by ICTP, GGI, MIAPP (funded by the Deutsche Forschungsgemeinschaft, German Research Foundation, under Germany's Excellence Strategy – EXC-2094 – 390783311.) and Swiss National Science Foundation (via project The Non-Gaussian Universe and Cosmological Symmetries, project number: 200020-178787). AL was supported in part by the Black Hole Initiative at Harvard University, which is funded by grants from JTF and GBMF.

%%%%%%%%%%%%%%%%%%%%%%%%%%%%%%%%%%%%%%%%%%%%%%%%%%

%\section*{Acknowledgements}

%The Acknowledgements section is not numbered. Here you can thank helpful colleagues, acknowledge funding agencies, telescopes and facilities used etc. Try to keep it short.

%%%%%%%%%%%%%%%%%%%%%%%%%%%%%%%%%%%%%%%%%%%%%%%%%%



\bibliographystyle{mnras}
\bibliography{SMFP} % if your bibtex file is called example.bib

%%%%%%%%%%%%%%%%%%%%%%%%%%%%%%%%%%%%%%%%%%%%%%%%%%

%%%%%%%%%%%%%%%%% APPENDICES %%%%%%%%%%%%%%%%%%%%%

%\appendix


%If you want to present additional material which would interrupt the flow of the main paper, it can be placed in an Appendix which appears after the list of references.

%%%%%%%%%%%%%%%%%%%%%%%%%%%%%%%%%%%%%%%%%%%%%%%%%%

% Don't change these lines
\bsp	% typesetting comment
\label{lastpage}
\end{document}